\documentclass[aps,preprint]{revtex4}%
\usepackage{amsfonts}
\usepackage{amsmath}
\usepackage{amssymb}
\usepackage{graphicx}%
\begin{document}
\title[Short title for running header]{Spin Dynamics of t-J Model on Triangular Lattice}
\author{Tao Li and Yong-Jin Jiang}
\affiliation{Center for Advanced Study, Tsinghua University, Beijing 100084, P.R.China}

\begin{abstract}
We study the spin dynamics of t-J model on triangular lattice in the
Slave-Boson-RPA scheme in light of the newly discovered superconductor
Na$_{x}$CoO$_{2}$. We find resonant peak in the dynamic spin susceptibility in
the $d+id^{^{\prime}}$-wave superconducting state for both hole and electron
doping in large doping range. We find the geometrical frustration inherent of
the triangular lattice provide us a unique opportunity to discriminate the
SO(5) and RPA-like intepretation of the origin of the resonant peak.

\end{abstract}
\volumeyear{year}
\volumenumber{number}
\issuenumber{number}
\eid{identifier}
\startpage{1}
\endpage{10}
\maketitle

The newly discovered superconductor Na$_{x}$CoO$_{2}$ has aroused many
interests\cite{1,2,3,4,5}. It is believed that its low energy physics is
analogous to high temperature superconductor and is another example of doped
2D Mott insulator. Another interesting aspect of this material is that it has
a triangular lattice and is geometrically frustrated and non-bipartite. An
effective model for this material is the t-J model on a triangular
lattice.\bigskip\ Historically, triangular lattice was used by Anderson to
introduce the concept of resonant valence bond(RVB) state which is now widely
used in the study of high temperature superconductivity, with the hope that
the geometrical frustration can help to stabilize the RVB state over the
ordered state.\bigskip\ Hence it is believed that the study of Na$_{x}%
$CoO$_{2}$ superconductivity may shed new light on the study of high
temperature superconductivity.

The most prominent feature of a doped Mott insulator is its spin dynamics,
since its charges are largely frozen by the strong correlation effect at low
energy. In high temperature superconductor, the spin dynamics plays a very
important role in our understanding of its physics. The most remarkable thing
in the spin dynamics of a high temperature superconductor is the appearance of
the so called resonant neutron peak in the superconducting state\cite{6}. The
origin of this remarkable phenomena (the dramatic change of spin dynamics
accompanying the appearance of coherence in charge motion) is still not well
understood. Two frequently discussed scenarios are a anti-bonding collective
mode in the particle-particle channel related to a SO(5) symmetry between
antiferromagnetism and d-wave superconductivity\cite{7,8}, and a particle-hole
bonding collective mode that manifests the internal structure of the Cooper
pair as in the RPA-like theory\cite{9,10,11}.

In this paper, we study the spin dynamics of t-J model on the triangular
lattice. We work in the Slave-Boson-RPA scheme and find resonant neutron peak
in the spin dynamics of the $d+id^{^{\prime}}$-wave superconducting state.
This is expected since the resonant neutron peak is a manifestation of the
internal structure of the Cooper pairs\cite{11} and it should be visible when
the measuring wavelength is comparable to the size of the pair(for
$d+id^{^{\prime}}$-wave pairing  this length scale is just the lattice size
near half filling). Since Cooper pair on the triangular lattice differ in
detail from that on the square lattice, we expect the resonant peak on the
triangular lattice will be different from that of high temperature superconductors.

The geometric frustration of the triangular lattice also provide us the
opportunity to discriminate the SO(5) and the RPA explanation of the origin of
the resonant peak. For the SO(5) explanation to apply, the two particle
continuum must collapse at some point in the Brillouin Zone in order to make
room for antibonding collective mode in the particle-particle channel to be
well defined at low energy. For a non-frustrated bipartite lattice, this
requirement is automatically met at $\left(  \pi,\pi\right)  $ point in the
Brillouin Zone(this property is used to show the SO(4) symmetry of the
negative-U Hubbard model, which is the origin of the SO(5) idea\cite{12}). For
the non-bipartite triangular lattice, the two particle continuum never
collapse and SO(5) explanation dose not apply(it is easy to prove that the
physical requirement of collapse of the two particle continuum is equivalent
to bipartite of the lattice for a general tight binding Hamiltonian\cite{13}).
Hence if we do detect resonant peak on triangular lattice, it must be caused
by some mechanism other than SO(5). In other words, the SO(5) idea is not
appropriate for a non-bipartite system.

Our starting point is the t-J model on a triangular lattice%

\[
H=-t%
{\displaystyle\sum\limits_{<i,j>,\sigma}}
(\hat{c}_{i,\sigma}^{\dagger}\hat{c}_{j,\sigma}+h.c.)+J%
{\displaystyle\sum\limits_{<i,j>}}
S_{i}\cdot S_{j}
\]

here $\hat{c}_{i,\sigma}$ is the constrained Fermion operator ($%
{\displaystyle\sum\limits_{\sigma}}
\hat{c}_{i,\sigma}^{\dagger}\hat{c}_{i,\sigma}\leq1$). To treat such
constraint we work in the Slave Boson scheme. This scheme is convenient for
the discussion of singlet superconductivity (for triplet pairing this scheme
is not convenient since the exchange term is repulsive in the triplet
channel). According to this scheme, we write $\hat{c}_{i,\sigma}=f_{i,\sigma
}b_{i}^{\dagger}$, in which $f_{i,\sigma}$ is the chargeless Fermionic spinon
operator and $b_{i}^{\dagger}$ is the spinless bosonic holon operator. Then,
we make the usual RVB\ mean-field decoupling (assuming uniform solution) and
arrive at the mean field Hamiltonian in the spinon sector (for simplicity we
only consider zero temperature and the holon is condensed in this case),%

\[
H_{MF}=-tx\sum_{<i,j>,\sigma}(f_{i,\sigma}^{\dagger}f_{j,\sigma}%
+h.c.)-\frac{3J}{8}\sum_{<i,j>}(\chi_{i,j}f_{i,\sigma}^{\dagger}f_{j,\sigma
}+\Delta_{i,j}f_{i,\uparrow}^{\dagger}f_{j,\downarrow}^{\dagger}+h.c.)-\mu
\sum_{i,\sigma}f_{i,\sigma}f_{i,\sigma}
\]

here $x$ is the holon concentration, $\chi_{i,j}=\left\langle f_{j,\uparrow
}^{\dagger}f_{i,\uparrow}+f_{j,\downarrow}^{\dagger}f_{i,\downarrow
}\right\rangle $ and $\Delta_{i,j}=\left\langle f_{i,\downarrow}f_{j,\uparrow
}+f_{j,\downarrow}f_{i,\uparrow}\right\rangle $ are RVB order parameter, $\mu$
is the chemical potential constraining the total Fermion number. Symmetry
consideration (rotational symmetry and the even parity of the singlet pairing)
 shows that the pairing order parameter $\Delta_{i,j}$ has only two
solution: a $d+id^{^{\prime}}$-wave solution and a extended s-wave solution
(the same conclusion is also reached by unconstrained Hartree-Fock
search\cite{4}). Since the s-wave solution is always higher in free energy, we
neglect it. The $d+id^{^{\prime}}$-wave solution is a analog of the d-wave
solution on square lattice. That is, the pairing extend over only nearest
neighbors, and a $\frac{2\pi}{z}$ rotation give rise to a $\frac{4\pi}{z}$ (or
$-\frac{4\pi}{z}$) phase change, where $z$ is the coordinate number of the
corresponding lattice($z=4$ for square lattice and $z=6$ for triangular
lattice). In square lattice case, such a pairing pattern is responsible for
the appearance of the resonant peak at $(\pi,\pi)$. Similarly, we expect the
triangular lattice can also support resonant peak but at a somewhat different
momentum region in the Brillouin Zone.

We first solve the self-consistent equation for the RVB order parameter. In
the triangular lattice, the particle-hole symmetry is broken and we solve for
both hole doping ($t>0$) and electron doping ($t<0$) case. The result is shown
in Figure 1. Here we present the result only for $\left\vert t\right\vert
/J=3$ (which is the value for the high temperature superconductors) since the
exact value of this parameter is still unknown\cite{4,5}. In general, the
doping range in which the $d+id^{^{\prime}}$-wave pairing order exists enlarge
with decreasing $\left\vert t\right\vert /J$ value, with the qualitative
features unchanged.

The calculation of the spin dynamics is straightforward. We first calculate
the mean field susceptibility according to the BCS formula:%

\[
\chi_{_{0}}(q,\omega)=\frac{1}{4}\sum_{k}(1-\frac{\xi_{k}\xi_{q-k}%
+\operatorname{Re}(\Delta_{k}\Delta_{q-k}^{\ast})}{E_{k}E_{q-k}})(\frac
{1}{\omega+E_{k}+E_{q-k}+i\delta}-\frac{1}{\omega-E_{k}-E_{q-k}+i\delta})
\]

here $\xi_{k}=-2(tx+\frac{3J}{8}\chi)(\cos(k_{x})+2\cos(\frac{k_{x}}{2}%
)\cos(\frac{\sqrt{3}k_{y}}{2}))-\mu$ is dispersion of the nearest-neighbouring
hopping Hamiltonian on the triangular lattice, $\Delta_{k}=\frac{3\Delta}%
{4}((\cos(k_{x})-\cos(\frac{k_{x}}{2})\cos(\frac{\sqrt{3}k_{y}}{2}))+i\sqrt
{3}\sin(\frac{k_{x}}{2})\sin(\frac{\sqrt{3}k_{y}}{2}))$ is the $d+id^{^{\prime
}}$-wave pairing order parameter which is complex(time reversal symmetry is
broken and there is staggered current loop flowing in the superconducting
state around every elementary triangular plaqutte). $E_{k}=\sqrt{\xi_{k}%
^{2}+\left\vert \Delta_{k}\right\vert ^{2}}$ is the energy of the BCS
quasiparticle. $\delta$ is its inverse life time. $(1-\frac{\xi_{k}\xi
_{q-k}+\operatorname{Re}(\Delta_{k}\Delta_{q-k}^{\ast})}{E_{k}E_{q-k}})$ is
the BCS coherence factor. In the square lattice case, since $\Delta
_{k}=-\Delta_{q-k}$ for $q=(\pi,\pi)$, the BCS coherence factor reach their
maximum value at $(\pi,\pi)$. However, for the non-bipartite triangular
lattice, there is no such special momentum.

The bare susceptibility should be renormalized by the RPA correction. The
correction is given by%

\[
\chi(q,\omega)=\frac{\chi_{0}(q,\omega)}{1-J(q)\chi_{0}(q,\omega)}
\]

here $J(q)=J(\cos(q_{x})+2\cos(\frac{q_{x}}{2})\cos(\frac{\sqrt{3}q_{y}}{2}))$
is the Fourier transform of the exchange interaction on the triangular
lattice. In general, the momentum dependence of the dynamic spin
susceptibility in the RPA-like theory is determined by the Fermi surface,
pairing gap(which together determine the BCS coherence factor that embody the
internal structure of Cooper pair) and the RPA renormalization factor $J(q)$
(which determine the classical spin ordering tendency). For d-wave pairing on
square lattice near half filling, both the BCS coherence factor and the RPA
renormalization factor are largest at $(\pi,\pi)$. For triangular lattice,
$J(q)$ reach its maximum at $(\frac{4\pi}{3},0)$ and symmetry related
points($(\pm\frac{4\pi}{3},0),(\pm\frac{2\pi}{3},\pm\frac{2\pi}{\sqrt{3}})$).
At the same time, although $\Delta_{k}=-\Delta_{q-k}$ dose not hold true
generally for any given $q$, calculation shows on average this condition is
least violated at the same set of momentum as $J(q)$ dose. Hence we expect to
see resonant peak around these set of momentum.

For square lattice, the RVB state is unstable with respect to RPA correction
in a large doping range. Two facts are responsible for this instability.
First, the system is bipartite and the Fermi surface is nearly nested. Second,
the exchange term is unfrustrated on square lattice. To cure this problem,
some author introduced a artificial reduction factor for the RPA
correction\cite{10}. For the non-bipartite triangular lattice, the single
particle dispersion is almost isotropic around the Fermi surface close to half
filling. At the same time, the exchange interaction is also frustrated. Hence
we expect RPA correction to give much weaker modification to the phase
diagram. Calculation shows for $\left\vert t\right\vert /J=3$ the ordering
instability occurs at about $2\%$ doping for hole doping and at about $2.5\%$
for electron doping (the ordering instability is indicated by the divergence
of the static spin susceptibility. For both hole and electron doping the
ordering instability occur near $(\frac{4\pi}{3},0)$ and symmetry related
momentum which correspond to the classical 3-sublattice $120^{0}$ spin order).
Hence, the RVB state is much more stable on triangular lattice than on square
lattice due to the geometrical frustration. For this reason we do not need to
introduce the reduction factor for the RPA correction.

We now present the result for the dynamic spin susceptibility. For $\left\vert
t\right\vert /J=3$, we find resonant peak below about $12\%$ doping for hole
doping and below about $7.5\%$ doping for electron doping. Hence there is a
large doping range in which the resonant peak can be detected, especially for
hole doping. The doping range in which resonant peak exists
increases(decreases) with decreasing (increasing) $\left\vert t\right\vert
/J$. Figure.2 show the energy dependence of the spin susceptibility at 5\%
doping. The momentum is taken at $(\frac{4\pi}{3},0)$. The dispersion of the
resonant peak is somewhat complicate. In Figure 3, we show momentum scans of
the dynamic spin susceptibility at various energy for 7\% hole doping. The
result can be summarized as follow (1)there is an optimal energy at which the
spin susceptibility reach its maximum. Above this energy, the resonant peak
appears at symmetric position on the $k_{x}$ axis (and symmetry related
momentum in the Brillouin Zone). Below the optimal energy, the resonant peak
split into two and disperse away from the $k_{x}$ axis. (2)The range of energy
below the optimal energy in which resonant peak exists shrinks with decreasing
doping and vanishes when ordering instability occurs. Hence the ordering
instability always occurs on the $k_{x}$ axis. (3)There is no qualitative
difference between electron doping and hole doping.

In the study of high temperature superconductivity, the origin of the resonant
peak and its role in the low energy physics are hotly debated. In the RPA-like
perspective, the resonant peak is a collective mode in the particle-hole
channel whose long life time is understood in kinematic manner. The existence
of this peak is a consequence of the superconducting pairing and its property
depends on the detailed form of the pairing. At the same time, this resonant
peak can couple to many other degree of freedom and make it visible in many
other experiments such as ARPES, STM, and optical conductivity\cite{14,15,16}.
Another interesting perspective of the resonant peak is the SO(5) theory of
high temperature superconductivity. In this perspective, the resonant peak is
a Goldstone mode relating the d-wave superconducting order and the
antiferromagnetic order. The stability of the mode is guaranteed by the SO(5)
symmetry. In fact, the existence of this particle-particle-channel
anti-bonding(since it is a spin triplet) collective mode is totally
independent of superconductivity in the SO(5) theory. Superconductivity only
make it visible in the spin channel and in this way providing a driving force
for superconductivity itself. Since it is a Goldstone mode related with a
symmetry, it is not expected that the mode to couple strongly with other
degree of freedom. However, even with these big differences, it is still hard
to decide which perspective is more appropriate for the high-Tc physics. Here
we argue that the study on the non-bipartite triangular lattice may provide
the opportunity to discriminate the two perspectives. This is because a
bipartite lattice is a prerequisite for the existence of SO(5) symmetry and of
well defined anti-bonding particle-particle collective mode at low energy.
Historically, the SO(5) symmetry and the particle-particle collective mode are
extensions of the SO(4) symmetry and the corresponding $\eta$ pair mode
relating CDW order and s-wave superconducting order in the negative U Hubbard
model\cite{12}. In that case, the existence of SO(4) symmetry and the $\eta
$\ pair rely on the bipartite property of the lattice ($\xi_{k}+\xi_{(\pi
,\pi....)-k}=const$). That is , the two particle continuum must collapse at
$(\pi,\pi....)$. More directly, it is this collapse of the two particle
continuum that makes a well defined antibonding particle-particle collective
mode possible at low energy(as we have mentioned, bipartite of the lattice is
equivalent to collapse of the two particle continuum at some momentum in the
Brillouin Zone). For the geometrically frustrated triangular lattice, the
collapse of two particle continuum never occur and there is no room for
antibonding particle-particle collective mode to be well defined at low
energy. To see this, we plot the momentum dependence of the band width of the
two particle continuum for both the square and the triangular lattice in
Figure. 4. From this plot, we find that the bandwidth of the two particle
continuum only vary mildly in the Brillouin Zone and never collapse to a point
for the triangular lattice. According to our discussion, this means the SO(5)
idea does not apply for such a geometrically frustrated system. Hence if we do
detect resonant peak on a triangular lattice system, its origin must not be
SO(5). Put it in another way, the SO(5) theory can only be defined on a
bipartite lattice.

\bigskip The authors would like to thank members of the HTS group at CASTU for discussion.

\begin{center}
\newpage

FIGURES
\end{center}

FIG. 1. Doping dependence of the $d+id^{^{\prime}}$-wave pairing order
parameter for both electron and hole doping.

\medskip

FIG. 2. Imaginary part of the dynamic spin susceptibility at $(\frac{4\pi}%
{3},0)$ for 5
lines are the bare and the RPA-corrected results.

\medskip

FIG. 3. Momentum scans of the dynamic spin susceptibility at various energy
for 7\% hole doping. (a) $\omega=0.39J$, (b) $\omega=0.40J$, (c)
$\omega=0.405J$, (d) $\omega=0.41J$, (e) $\omega=0.415J$, (f) $\omega=0.42J$,
(g) $\omega=0.425J$, (h) $\omega=0.43J$. The maximum of the dynamic spin
susceptibility occurs at $\omega=0.415J$ at this doping rate.

\medskip

FIG. 4. Momentum dependence of reduced band width of the two particle
continuum for (a)the bipartite square lattice, (b)the non-bipartite triangular lattice.


\begin{thebibliography}{99}
\bibitem {13}For a general tight binding model $%
{\displaystyle\sum\limits_{i,\delta}}
t_{\delta}c_{i}^{\dagger}c_{i+\delta}$ (assuming reflection symmetry hold),
the requirement of the collapse of the two particle continuum at total
momentum $q$ ($\xi_{k}+\xi_{q-k}=\xi_{k}+\xi_{k-q}=%
{\displaystyle\sum\limits_{\delta}}
t_{\delta}(1+e^{-iq\delta})e^{ik\delta}=const.$) implies the hopping matrix
element $t_{\delta}$ is nonzero only for those neighbours that $e^{-iq\delta
}=-1$ and the hopping matrix element between two such neighbours is zero.
Hence the lattice is bipartite. The reverse is also true. For a bipartite
lattice, collapse of the two particle continuum occurs at all momentum $q$
such that $e^{-iq\delta}=-1$, where $\delta$ is a arbitrary vector connecting
sites on the two sublattices.
\end{thebibliography}
\end{document}